\documentclass[journal=jacsat,manuscript=article]{achemso}
\usepackage{hyperref}
\usepackage{graphicx} 
\usepackage{microtype}
\usepackage{upgreek}
\usepackage{wrapfig}

\usepackage[T1]{fontenc} 
\usepackage[version=3]{mhchem}

\def \MSE{Department of Materials Science and Engineering, National University of Singapore, 117575, Singapore}
\def \IFIM{Institute for Functional Intelligent Materials, National University of Singapore, 117544, Singapore}
\def \JAPAN{Research Center for Functional Materials, National Institute for Materials Science, Tsukuba 305-0044, Japan}
\def \JAPANN{International Center for Materials Nanoarchitectonics, National Institute for Materials Science, 305-0044, Japan}
 
\author{M.~Grzeszczyk}
\email{magda@nus.edu.sg}
\affiliation{\IFIM}
\author{K.~\,Vaklinova}
\affiliation{\IFIM}
\author{K.~\,Watanabe}
\affiliation{\JAPAN}
\author{T.~\,Taniguchi}
\affiliation{\JAPANN}
\author{K.~S.~\,Novoselov}
\affiliation{\IFIM}
\alsoaffiliation{\MSE}
\author{M.~\,Koperski}
\email{msemaci@nus.edu.sg}
\affiliation{\IFIM}
\alsoaffiliation{\MSE}

\title{Electrical excitation of carbon centers in hexagonal boron nitride with tuneable quantum efficiency.}

\begin{document}

\begin{abstract}

Defect centers in wide-band-gap crystals attracted considerable attention due to the realisations of qubits, sensors, or single photon emitters at room temperature. The family of these centers is constantly growing, including well-known examples such as nitrogen-vacancy centers in diamond, silicon-vacancy in silicon carbide, chromium substitutions in aluminium oxide, and many others. Unfortunately, such defect centers embedded in highly insulating crystals have been notoriously difficult to excite electrically. Herewith, we present a realisation of insulating light-emitting diodes based on carbon centers in hexagonal boron nitride. The rational design of the vertical tunnelling devices via van der Waals technology enabled us to control the charge dynamics related to non-radiative tunelling, defect-to-band electroluminescence, and intradefect electroluminescence. The fundamental understanding of the tunnelling events enabled us to achieve high efficiency of electrical excitation, which exceeded by a few orders of magnitude the efficiency of optical excitation in the sub-band-gap regime. A combination of a Stark effect and screening by band electrons provide a control knob for tuning the energy of emission. With this work, we solve an outstanding problem of creating electrically driven devices realised with defect centers in wide-band-gap crystals, which are relevant in the domain of optoelectronics, telecommunication, computation, or sensing.

\end{abstract}


\section{Introduction}

Functionalities of crystals at an atomic scale can be activated through the systematic, controllable, and precise formation of defect centers. From such a perspective, wide-band gap materials are particularly attractive \cite{heremans2016control,castelletto2020silicon,diler2020coherent}. The crystallographic modifications in (sub--) nano--regime may lead to formations of spin centres that are isolated from the fundamental electronic bands, becoming simple and pure quantum systems \cite{NV_EPR, spin_manip, spin_manip2, coherence_NV}. When properly mastered, defect centers act as qubits \cite{defect_comp, NV_qubit}, single photon emitters \cite{hBN_quantum_emission, hBN_quantum_emission2, koperski2021towards, koperski2017optical, Koperski2018_hBN} \cite{quantum_photonics}, sensors of pressure \cite{ruby_pressure,gottscholl2021spin}, magnetic fields \cite{scanning, NV_sensing, Quantum_sensing}, electric fields\cite{Efield_sensing}, thermal conductivity \cite{therm_cond_image}, or dielectric constant\cite{singh2018engineering} at the ultimate limit of miniaturization. They also activate the optical response of materials, enabling the realization of lasers \cite{hBN_laser} and photodetectors \cite{defect_photodetector}. The defect levels located deep within the band gap \cite{hBN_deep_level_defect} enable operation at room temperature, making the research on such centres highly technological.  However, all these functionalities suffer from a significant bottleneck resulting from the insulating character of the host crystals. Large internal resistance prevents in most cases the development of electrically driven devices. \cite{song2021deep,moon2022hexagonal} Herewith, we address this problem by the creation and characterization of electroluminescent tunneling diodes operating in near ultraviolet, visible, and near-infrared spectral ranges utilizing hexagonal boron nitride (hBN) as a dielectric barrier material and an optically active medium. The post-growth carbon doping \cite{hBN_carbon_doping, koperski2020midgap} combined with the engineering of the device architecture enables the electrical injection of electron-hole pairs with tunable quantum efficiency related to the intra--defect and defect-to-band optical transitions.

The advantage of hBN over other materials originates from the two-dimensional honeycomb crystal structure. A broken symmetry between the boron and nitrogen triangular sub-lattices opens a band gap of the order of 6~eV \cite{watanabe2011hexagonal, watanabe2004direct, cassabois2016hexagonal, caldwell2019photonics, schue2019bright,paleari2019exciton,cannuccia2019theory}, establishing hBN as a "white graphene" system. Due to matching crystal structures, hBN/graphene (Gr) devices feature inert, atomically flat, and homogeneous surfaces. High-quality interfaces in the nanoscale enable the wavefunctions of electrons in graphene electrodes to penetrate the hBN dielectric barrier hosting mid-gap defect levels. The spatial proximity of electronic states enables the rational design of devices with controllable tunneling processes, including those that lead to light-emitting events.

hBN has taken a central role in many areas of research due to its unique and diverse characteristics, including a low dielectric constant\cite{hong2020ultralow}, high thermal conductivity\cite{zheng2016high,cai2019high}, and chemical inertness\cite{lei2021low,li2016atomically}. hBN has become a key ingredient in van der Waals technology, acting as a substrate for assembly and growth \cite{hBN_substrate}, a dielectric environment for the engineering of local electrostatic screening \cite{hBN_dielectric_env}, tunnelling barrier in devices \cite{hBN_tunnelling}, ultraviolet lasers \cite{hBN_laser}, and a source of single photons \cite{hBN_quantum_emission, hBN_quantum_emission2} for optoelectronic applications. Thin (less than 100 atoms) light-emitting diodes (LEDs) with hBN as the optically active medium create a platform for electrical excitations of well-defined defect centers in a wide band-gap system.

\section{Results and discussion}

\subsection{Samples}

\begin{figure}
    \centering
    \includegraphics[width=\linewidth]{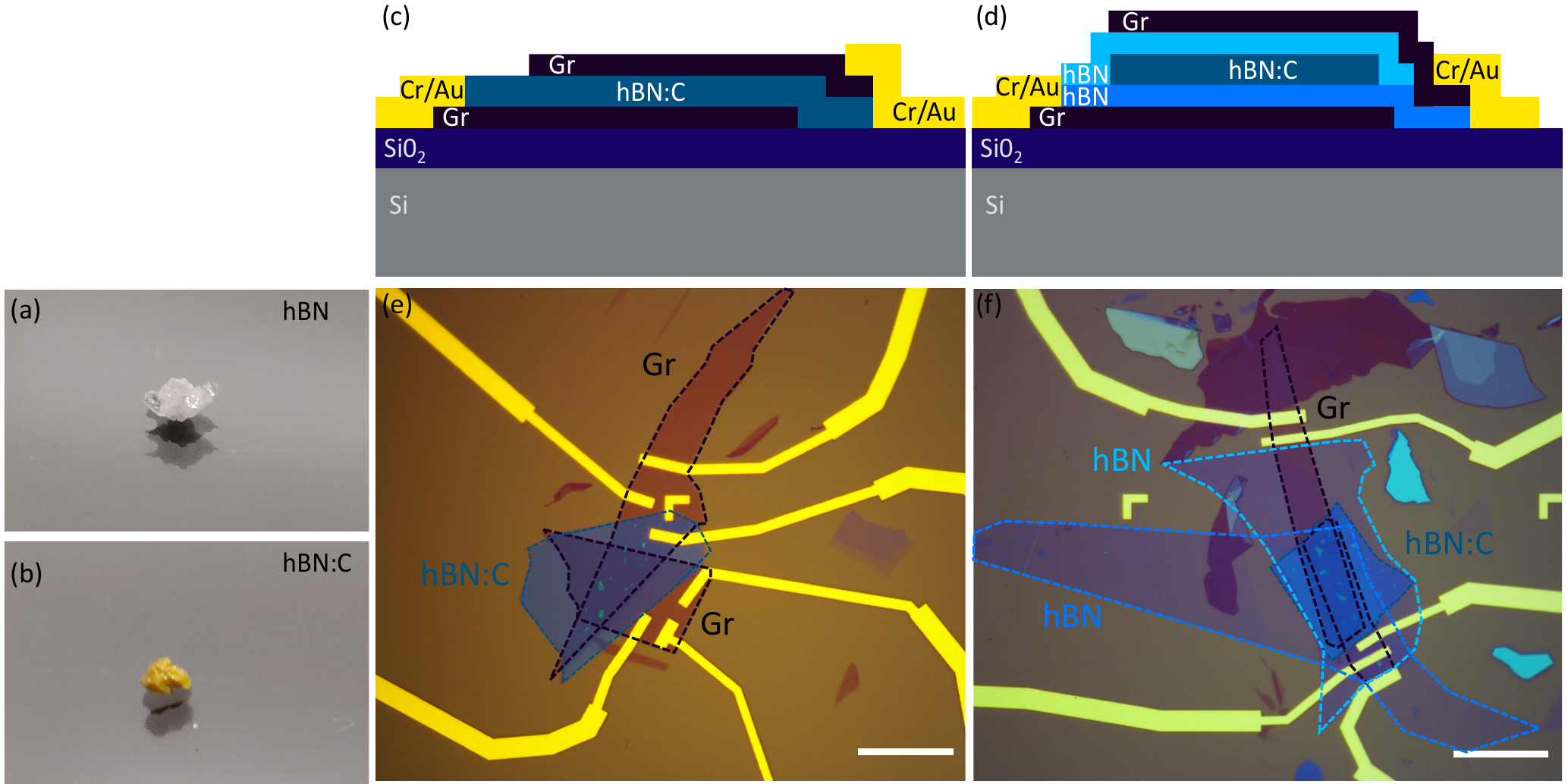}
\caption{\textbf{The architecture of the vertical tunnelling devices with hBN:C as an energy barrier}. The optical images of the hBN (a) and hBN:C bulk crystals (b). Schematic representation of the light emitting diode structures with the layer sequence Gr/hBN:C/Gr (c) and Gr/hBN/hBN:C/hBN/Gr (d). The devices were created by stacking mechanically exfoliated graphene, hBN, and hBN:C layers. The optical images of example devices with Gr/hBN:C/Gr and Gr/hBN/hBN:C/hBN/Gr architectures are presented in (e) and (f), respectively. The thickness of hBN:C was $\sim$20~nm, while the hBN barriers were approximately 3-5~nm thick. The scale bar corresponds to 10~$\upmu$m.}
    \label{fig:samples}
\end{figure}

The creation of the hBN-based LEDs was enabled through the introduction of controllable post-growth carbon-doping \cite{hBN_carbon_doping}. It is critical that the crystallographic quality of the hBN is high for the thin films to fulfill a role of a well-defined tunnelling barrier \cite{hBN_tunnelling}. To that end, we grew the bulk hBN crystals via high pressure, temperature-gradient method \cite{hBN_synth}. From a single growth, we annealed a fraction of the crystals in a graphite furnace at the temperature of 2000$^{\circ}$C for a period of time ranging from 1~h to 5~h to achieve varied concentrations of carbon impurities. The carbon-doped hBN (hBN:C) becomes yellow to blackish, depending on the annealing time, contrary to the pristine transparent hBN [for comparison of optical images see Fig.~\ref{fig:samples}(a)-(b)].

The carbon doping procedure activates multiple resonances in the photoluminescence (PL) spectra in the ultraviolet, visible, and near-infrared spectral regions \cite{koperski2020midgap}. The translation from optical to electrical excitation requires understanding and controlling the dynamics of the tunnelling processes. Electrically injected carriers onto the defect levels need to live long enough in the material to allow for radiative recombination. Therefore, we have fabricated two types of tunnelling devices which are schematically depicted in Fig.~\ref{fig:samples}(c)-(d). A Si/(290~nm) \ce{SiO2} wafer is used as the substrate for the vdW heterostructure with the following layer sequence: 1) Gr/hBN:C/Gr and  2) Gr/hBN/hBN:C/hBN/Gr. The optically active hBN:C layer was $\sim$20~nm thick and the additional hBN spacer was 3 - 5~nm thick for the devices discussed herein. Optical microscope images of the representatives of the two classes of devices are shown in Fig.~\ref{fig:samples}(e)-(f). Details about the device fabrication can be found in Supplementary Information (SI). 

\subsection{Results}


\begin{figure}[h!]
    \centering
    \includegraphics[width=\linewidth]{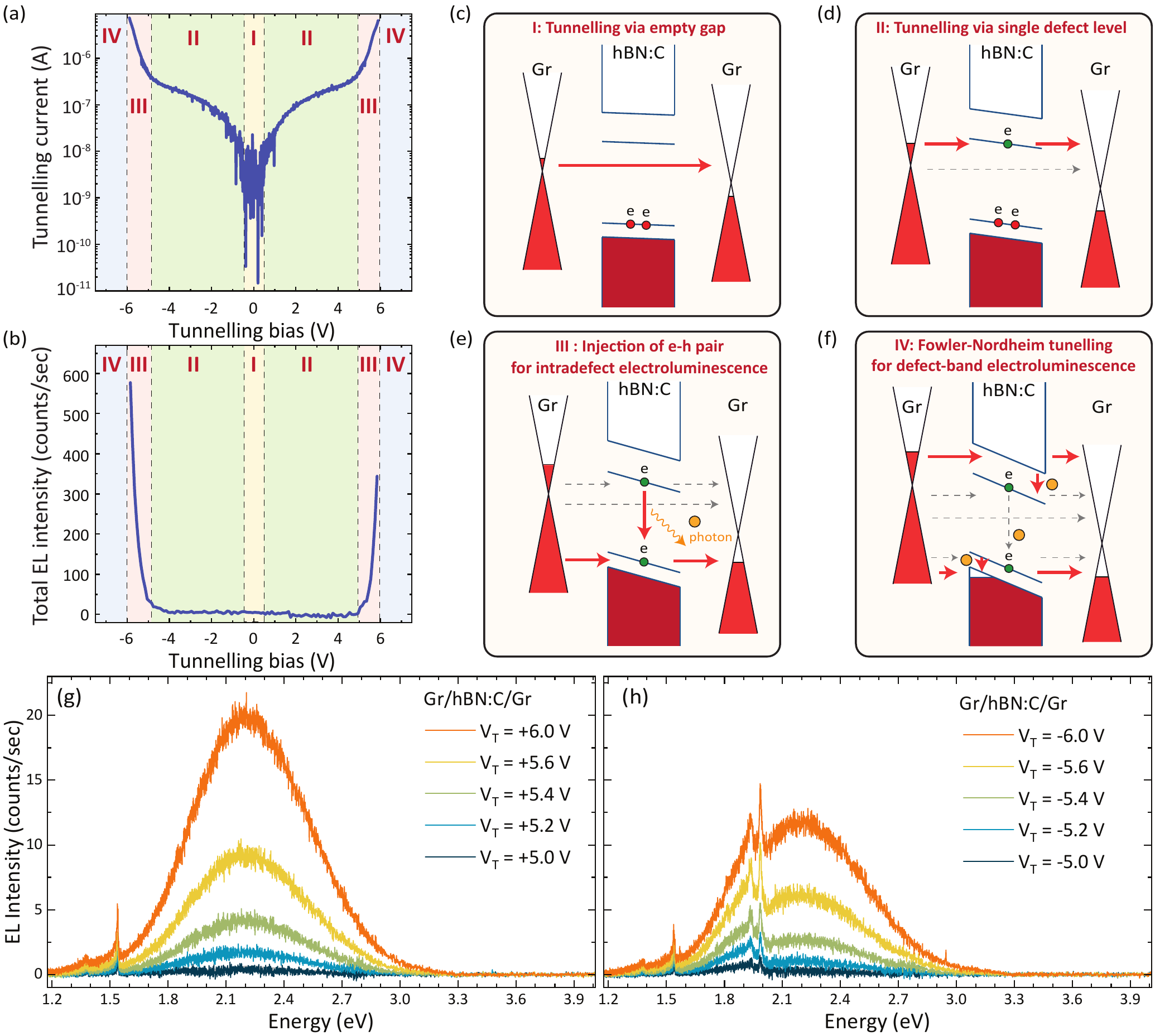}
\caption{\textbf{Electroluminescence from Gr/hBN:C/Gr devices.} Tunnelling \textit{IV} curve for a Gr/hBN:C/Gr (a) demonstrates activation of novel tunnelling paths. The device displays electroluminescence at a higher bias threshold, which can be seen from the dependence of the integrated light emission intensity on the tunnelling bias (b). The integration is done in the spectral region 1.2 -- 3.9~eV. Schematic depiction of the tunnelling processes based on the band structure of the heterostructure and the Fermi level alignment with bias is presented in various regimes: tunnelling via empty gap for small bias voltage (c), tunnelling via single defect level with an increased bias voltage (d), formation of electron-hole pairs activating intradefect electroluminescence (e), defect-to-band electroluminescence in the sup-band-gap bias regime (f). The electroluminescence spectra for different bias voltages are presented under the forward (g) and reverse (h) biasing direction. All the measurements were done at T = 5~K.}
    \label{fig:Dev1}
\end{figure}

We begin with the characterisation of the more straightforward Gr/hBN:C/Gr device, whose \textit{IV} characteristics are demonstrated in Fig.~\ref{fig:Dev1}(a). The zero-bias resistance yields 0.1~G$\Omega$, which enables insight into the activation of consecutive tunnelling paths upon the increase of the tunnelling bias. At the biases below 0.5~V, the current is given predominantly by the zero-bias resistance. We interpret this regime as a direct Gr-Gr tunnelling through an empty hBN band gap [Fig. ~\ref{fig:Dev1}(d)]. As the separation between the Gr electrodes is 20~nm, the overlaps of the electronic wave functions are expected to be negligible. The first onset of tunnelling is observable at the bias of about $\pm 0.5$~V. We associate this threshold with the tunnelling mediated by a defect state [Fig. ~\ref{fig:Dev1}(d)], when the Fermi level in a Gr electrode is brought in resonance with one of the hBN:C defect levels. In this condition, electroluminescence is not allowed due to the absence of an empty final state of the recombination process, if we assume tunnelling via a defect level below the conduction band. Indeed a second onset related to the increase of the slope of the \textit{IV} curve is observable at 5~V, which coincides with the detection of light emitted from the device.

The inspection of the electroluminescence spectra, which are demonstrated in Fig.~\ref{fig:Dev1}(g-h), grants further insight into the charge tunneling processes in this regime. Two types of contributions to these spectra can be identified: 1) narrow resonances characteristic of intradefect excitations and 2) broadband feature with a width of about 1~eV which can be expected from transitions between the mid-gap defect level and electronic sub-bands \cite{defect_to_band}. The intradefect electroluminescence requires the injection of a hole into the lower energy defect level. Consequently, a new tunnelling pathway opens up via this defect level [Fig.~\ref{fig:Dev1}(e)] accounting for the increase of the \textit{IV} slope. However, the activation of the broadband electroluminescence requires an alternative tunnelling process via the Fowler-Nordheim mechanism \cite{fowler1928electron, hattori2018determination}. Either an electron can be injected into the conduction band, [Fig. ~\ref{fig:Dev1}(f)] recombining with the underlying defect level, or a hole can be injected into the valence band enabling recombination of an electron occupying a defect level. The bias threshold for the intradefect and defect-to-band optical transitions depends on the electronic structure of the mid-gap levels in combination with the position of the Fermi level in graphene with respect to the band edges of hBN forming a triangular tunnelling barrier. In the Gr/hBN:C/Gr device the threshold for both types of electroluminescence coincide at about 5~V with the tunnelling current of 400~nA.

The analysis of this device architecture highlights the competition between the non-radiative Gr-Gr tunnelling, intradefect, and defect-to-band electroluminescence. From a technological perspective, it would be desirable to enhance the efficiency of the intradefect recombination pathway, while quenching the other two alternative tunnelling events. In a simple view, the electroluminescent processes should become more probable if the tunnelling time is increased, so that the charge carriers remain longer in the hBN:C material. We realise this notion by introducing additional pristine hBN barriers (2-3~nm thick) between the hBN:C optically active medium and the Gr electrodes.

\begin{figure}[h!]
    \centering
    \includegraphics[width=\linewidth]{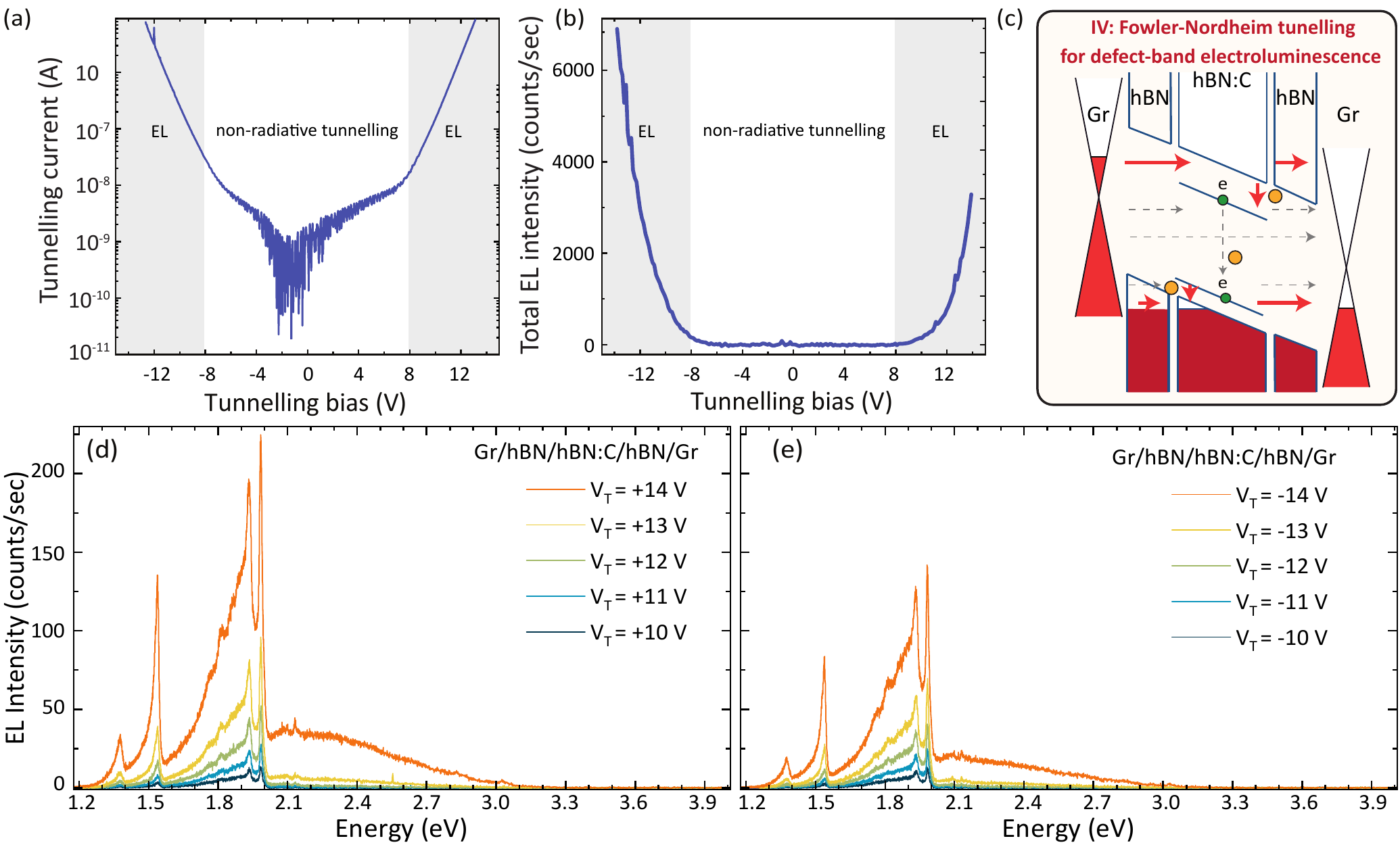}
\caption{\textbf{Electroluminescence from Gr/hBN/hBN:C/hBN/Gr devices.} Tunnelling \textit{IV} curves (a) and the integrated EL intensity dependence on the tunnelling bias (b) demonstrate the increase of the tunnelling and electroluminescence voltage threshold induced by additional pristine hBN barriers. Schematic depiction of the charge tunnelling processes in the sup-band-gap voltage regime enabling non-radiative Gr-Gr tunnelling (including tunnelling mediated by electronic bands via Fowler-Nordheim mechanism), defect-to-band electroluminescence and intradefect electroluminescence (c). The evolution of the electroluminescence spectra with bias voltage under forward (d) and reverse (e) direction demonstrate that the electroluminescence spectra are dominated by intradefect optical transitions. All the measurements were done at T = 5~K.}
    \label{fig:Dev2}
\end{figure}

The zero-bias resistance for the Gr/hBN/hBN:C/hBN/Gr device increases by more than two orders of magnitude to the value of $\sim$20~G$\Omega$. The consecutive tunnelling onsets are not conspicuous in the \textit{IV} curves due to large resistance [Fig. ~\ref{fig:Dev2}(a)]. Consequently, in this device, the onset of electroluminescence is driven by current rather than the voltage threshold. The light emission appears at the bias of 8~V with the current of 10~nA [Fig.~\ref{fig:Dev2}(b)]. Even though the voltage threshold increased, the electroluminescence requires 40 times smaller current to be activated. The reduction of current coincides with a qualitative modification of the electroluminescence spectra as demonstrated in Fig.~\ref{fig:Dev2}(d,e). In this device architecture, the intradefect transitions characterized by narrow linewidth dominate over the broadband contribution associated with defect-to-band transitions. The total emission intensity, integrated over the spectral region 1.2 - 3.3~eV defined by the convolution of the sensitivity of our optical setup with the optical response of the sample, also significantly increases. For example, for the current of $1~\upmu$A the integrated electroluminescence signal is about 20 times stronger for the device with additional barriers.

We interpret the barrier-induced modifications of the optoelectronic proprieties of our tunnelling devices in terms of the band alignment presented in a schematic in Fig.~\ref{fig:Dev2}(c). The efficiency of the individual tunnelling processes in this device is governed by the dynamics of the charge tunnelling and the radiative recombination in the sup-band-gap bias regime. The increased efficiency of light emission originates from enhanced tunneling time between the sub-conduction-band defect level and the graphene electrode. The electron remains longer in the hBN:C material, which increases the probability of the electron being injected into the sub-conduction-band level to follow the radiative recombination pathway. The probability of an electron escaping the hBN:C materials via non-radiative tunnelling depends on the character of the tunnelling process. Introduction of additional barriers should significantly increase the efficiency of processes leading to intradefect electroluminescnece. On the contrary, in the Fowler-Nodheim regime, the defect-to band electroluminescence should be weakly affected by additional barriers. The free electrons in the conduction band and the free holes in the valence band will relax along the large gradient of the electric field, enabling and easy escape to the opposite graphene electrodes. From a practival perspective, we have found that devices with 3~nm pristine hBN barriers are sufficient to obtain the dominant electroluminescence response from intradefect transitions. Notably, our devices oftentimes exhibit a slight asymmetry as seen in the \textit{IV} curves and electroluminescence response of intensity dependence on the tunnelling bias. The origin of the asymmetry may be attributed to a slight variation in the number of hBN layers and/or varied quality of the interfaces determined by the stacking process.

\begin{figure}[h!]
    \centering
    \includegraphics[width=\linewidth]{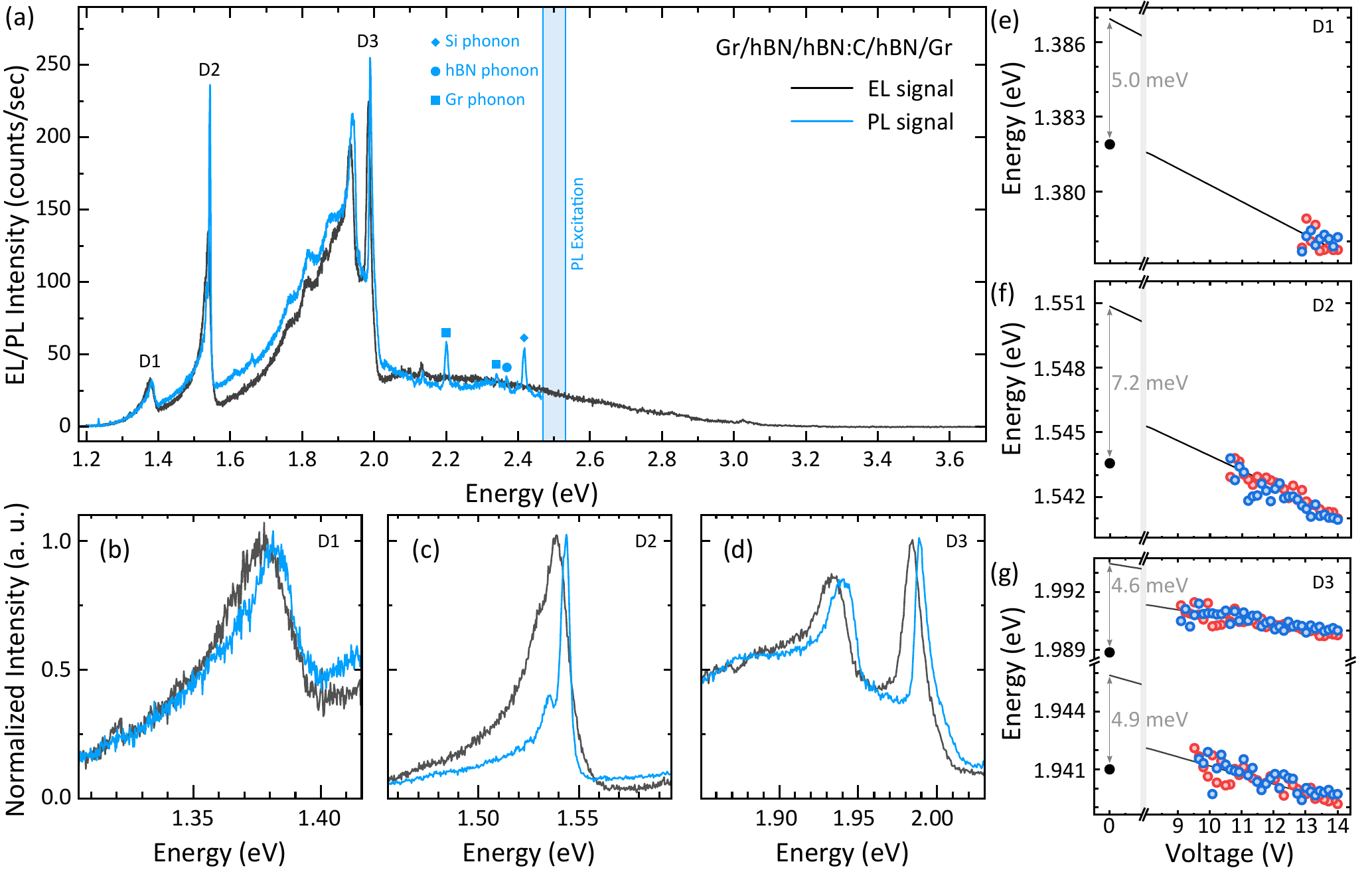}
\caption{\textbf{Electrical and optical excitation of carbon centers in hBN and the bias-driven control of the emission energy.} The comparison of the photoluminescence and electroluminescence spectra for the Gr/hBN/hBN:C/hBN/Gr device (a). The photoluminescence spectrum was collected with 2.56~eV laser excitation. The laser was focused to a spot of 1~$\upmu$m diameter with a power of 1~mW. The electroluminescence was measured with -14~V bias voltage, which corresponded to I$_\text{T}$ = -2~$\upmu$A. The photoluminescence and electroluminescence spectra were collected via the same microscope objective from a region of the sample with a size close to the diffraction limit for the adequate comparison of emission intensity. All the individual emission centers, D1 (b), D2 (c), and D3 (d), are present in both optically and electrically excited spectra. The energy of emission for defects D1 (e), D2 (f) and D3 (f) changes with the tunnelling bias in forward (red dots) and reverse (blue dots) directions. The extrapolated zero-bias energy is offset by a few meV from the energy of the resonance in the photoluminescence spectrum [black dot in (e)-(g)]. All the measurements were done at T = 5~K.}
    \label{fig:EL_PL}
\end{figure}

We can inspect more quantitatively the efficiency of the electrical excitations of the intradefect light emitters via a comparison between photoluminescence and electroluminescence spectroscopy. The electrically and optically excited light emission spectra from the same device are presented in Fig.~\ref{fig:EL_PL}(a). We identify three types of defects, labeled D1 - D3, that qualitatively produce the same signatures in the photoluminescence and electroluminescence response. Notably, comparable emission intensity is observed for the optical excitation at 2.56~eV with the power of 1~mW and for the electrical excitation with the current of 2~$\upmu$A. These values correspond to excitations with $2.4*10^{15}$ photons per second and $1.2*10^{13}$ electrons per second, making the photon-to-photon conversion over two orders of magnitude smaller than the electron-to-photon conversion. The significantly enhanced quantum efficiency for electrical excitation can be understood from geometrical and band structure considerations. The optical excitation is realised in the sub-band-gap regime, implying that the photons are absorbed resonantly at the defect sites, which constitute only a small fraction of the volume of the crystal (i.e., the host crystal is optically transparent). The \textit{IV} characteristics of our devices demonstrate that the charge tunnelling current in the sub-band-gap bias regime in the absence of available defect states is negligible (i.e., the host crystal is not conductive electrically). Consequently, in a simple view, the hBN crystal is transparent for photons but opaque for electrons in the energy scale smaller than the band gap, making the absorption of light very weak in the presence of a scarce distribution of defect centers.

Such quantitative comparison does not fully reflect the efficiency and advantages of electrical excitation. Firstly, we compared a local optical excitation with a laser spot focused to a spot of about 1~$\upmu$m diameter with bulk electrical excitation within the whole area of the device, which yields about 375~$\upmu$m$^2$. As the electroluminescence signal from the device is homogeneous (see Fig.~\ref{fig:maps} in the SI), therefore the current that flows through the area from which we detect emitted light corresponds to $2.5*10^{10}$ electrons per second, making the electrical excitation renormalised geometrically five orders of magnitude more efficient than the optical excitation. Secondly, we are comparing the electrical excitation above the band gap with the optical excitation below the band gap. This implies that a substantial portion of current flows through the conduction and valence band of hBN, potentially leading to interband electroluminescence, non-radiative Gr-Gr tunnelling, band-to-defect electroluminescence and relaxation of band carriers onto defect levels followed by intradefect electroluminescence. The efficiency of these processes can hardly be studied, as the optical setups cannot work efficiently from near-infrared to deeper ultraviolet spectral regions. For that reason, the above-band-gap optical excitation of hBN is not practical and rarely realised experimentally. Therefore the ability of electrical excitations opens a way to further explore the fundamental properties of hBN crystals, beyond the optical limitations.

Moreover, the electrical excitation enables control of the energy of the optical resonances, as illustrated in Fig. ~\ref{fig:EL_PL}(b-d). In the Fowler-Nordheim tunnelling regime we observe a redshift of the emission energy for three types of defects, D1 - D3, upon increasing the tunnelling bias. We interpret this observation as an effect of screening by electrons injected into to conduction band of hBN. Increased effective dielectric constant leads to the reduction of the energy of the intradefect optical transitions related to carbon and vacancy centers in hBN. In the below band gap regime for the Gr/hBN:C/Gr devices, the evolution of the energy of emission is qualitatively different, in agreement with a Stark effect \cite{Stark_NV, hBN_Stark} (see Fig.~S7 in the SI). Therefore, the energy tuneability is driven by a competition between the dielectric screening and Stark effect, with the dominant contribution determined by the tunnelling regime.


\section{Conclusions}

In summary, we demonstrated tunable electroluminescence in vertical tunneling devices with hBN acting as a dielectric barrier and an optically active medium. Through device architecture engineering, we activated multiple tunneling pathways, including those leading to defect-to-band and intradefect electroluminescence. Through comparative analysis with photoluminescence experiments, we have found that electrical excitation is several orders of magnitude more efficient than optical excitation realised in the commonly used sub-band gap regime. Our findings create a platform for electrical charge injections into wide band-gap materials, introducing electrical control degree of freedom to their many functionalities in the domain of optoelectronics, telecommunication, sensing, and spin physics. 

\begin{acknowledgement}
This project was supported by the Ministry of Education (Singapore) through the Research Centre of Excellence program (grant EDUN C‐33‐18‐279‐V12, I‐FIM), AcRF Tier 3 (MOE2018-T3-1-005). This research is supported by the Ministry of Education, Singapore, under its Academic Research Fund Tier 2 (MOE-T2EP50122-0012). This material is based upon work supported by the Air Force Office of Scientific Research and the Office of Naval Research Global under award number FA8655-21-1-7026. K.W. and T.T. acknowledge support from JSPS KAKENHI (Grant Numbers 19H05790, 20H00354 and 21H05233).
\end{acknowledgement}

\begin{suppinfo}

Supplementary Information available: Sample fabrication, experimental details, summary of other studied devices and their performance, additional data on dielectric breakdown and its effect on the EL, sample homogeneity, more detailed description of the emission behaviour versus applied bias voltage and description of alternative many-body picture of charging and electroluminescence mechanism. 
\end{suppinfo}

\bibliography{ref}

\pagebreak
\begin{center}
\textbf{\Large Supplementary Information for: \\ Electrical excitation of carbon centers in hexagonal boron nitride with tuneable quantum efficiency.}
\end{center}
\setcounter{equation}{0}
\setcounter{figure}{0}
\setcounter{table}{0}
\setcounter{page}{1}
\makeatletter
\renewcommand{\theequation}{S\arabic{equation}}
\renewcommand{\thefigure}{S\arabic{figure}}
\renewcommand{\bibnumfmt}[1]{[S#1]}
\renewcommand{\citenumfont}[1]{S#1}

\section{Sample fabrication}
Graphite, hBN, and hBN:C crystals were mechanically cleaved onto silicon wafers with 300~nm and 90~nm layers of \ce{SiO2}, respectively. Thin graphite films were selected to act as
electrodes for hBN:C of approximately 20~nm thickness, identified by optical contrast and atomic force microscopy. 
Using a polydimethylsiloxane/polycarbonate stamp the assemble of graphite/hBN/hBN:C/hBN or graphite/hBN:C/ was lifted from the Si/\ce{SiO2} wafer at 100$^{\circ}$C. Subsequently, the stacks were released onto the pre‐exfoliated graphite flake together with the polycarbonate film at 180$^{\circ}$C, which was washed away thereafter by using dichloromethane, acetone, and isopropanol to remove the polymer residues. The electrical contacts to the top and bottom graphite electrodes were patterned using electron‐beam lithography followed by evaporation of 5~nm‐Cr/60~nm‐Au layer and finalized by a lift‐off process.

\section{Experimental setup}

The optical spectra were measured in dry cryogenic systems with
a base temperature of 4.2~K. The sample was cooled down via thermal contact with a cold finger. The laser light was focalized at the surface of the device and the PL/EL signal was collected through an in‐situ objective with a 0.82 numerical aperture. The sample was mounted on a chip carrier positioned on a set of x/y piezo‐positioners that allow alignment, while the objective was fixed on a z piezo‐positioner enabling focalization of the laser light. The optical signal was dispersed by a 0.75~m spectrometer equipped with a 150~g/mm grating. The light was detected by liquid nitrogen-cooled charge‐coupled device camera. For PL excitation a \mbox{$\lambda$=488~nm} (2.56~eV) or \mbox{$\lambda$=514~nm} (2.41~eV) continuous wave (CW) laser diode was used. The excitation power focused on the sample was kept at 1~mW. The current-voltage (\textit{IV})  characteristics were measured with a Keithley 24XX source meter.

\newpage

\section{Summary of other studied devices and comments
on their performance.}

\begin{figure}[h!]
\centering
     \includegraphics[width=\linewidth]{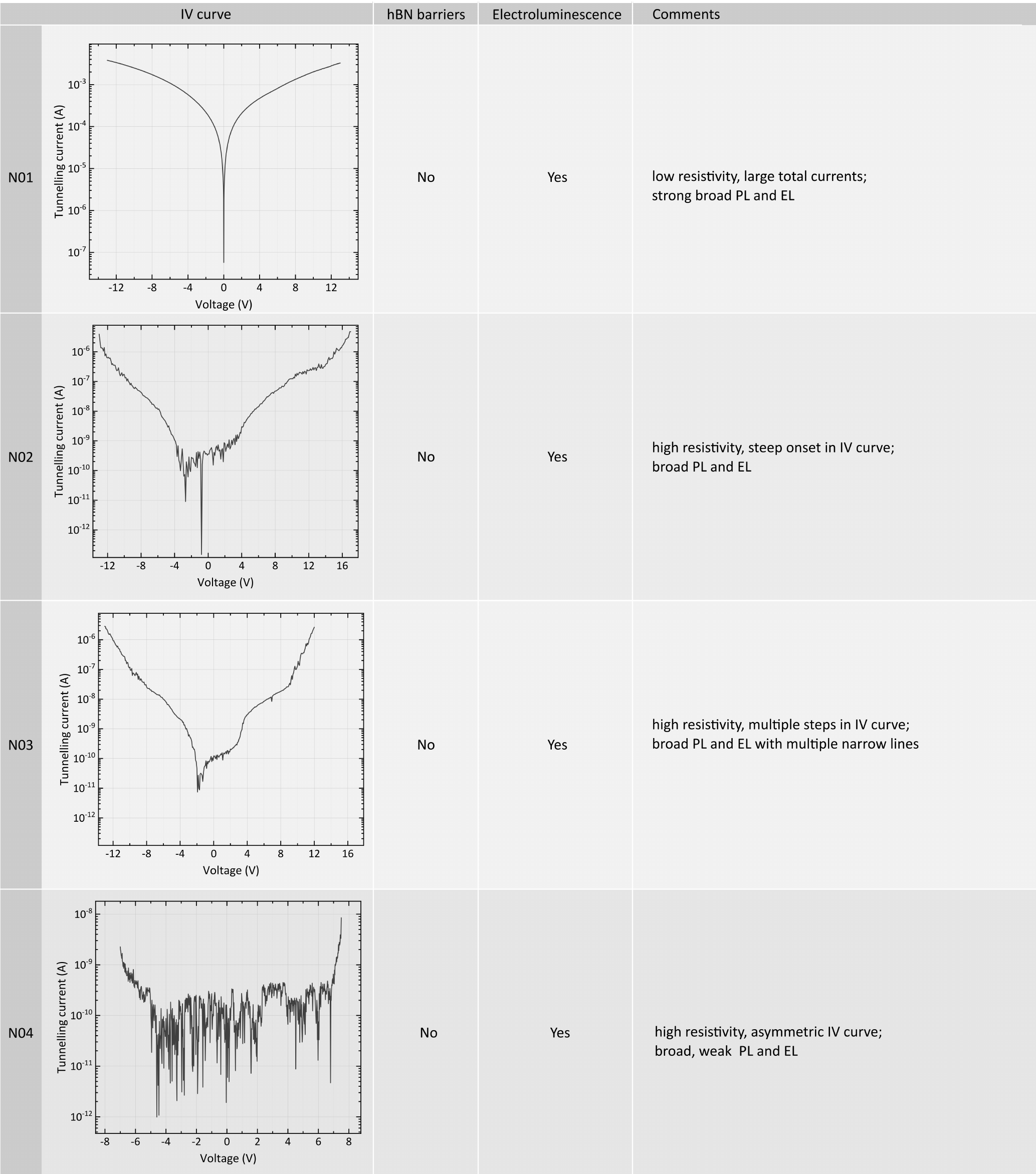}
    \caption{Current-voltage characteristics and additional information for all studied devices - part I.}
    \label{fig:summary1}
\end{figure} 

\newpage

 \begin{figure}[h!]
\centering
     \includegraphics[width=\linewidth]{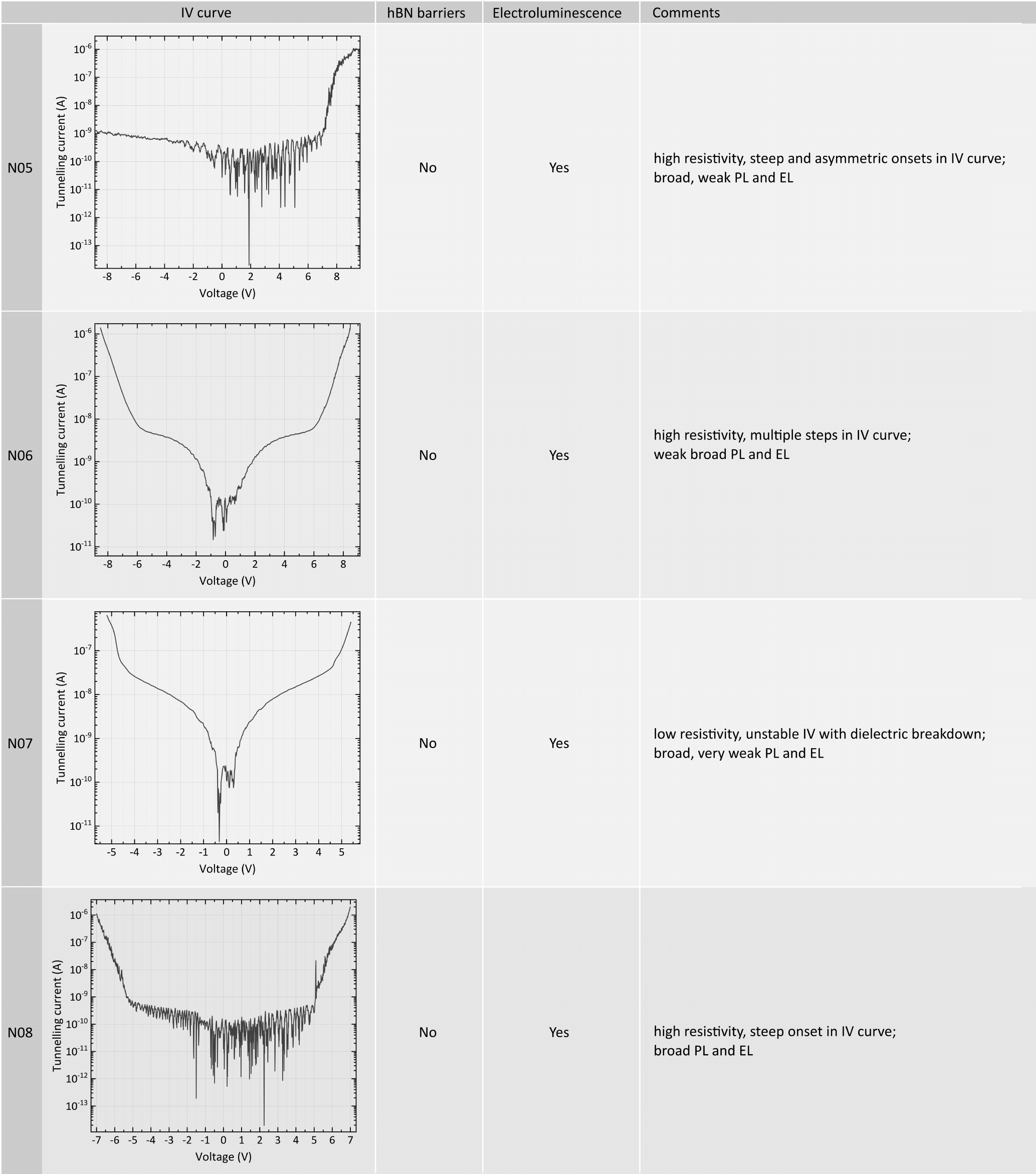}
    \caption{Current-voltage characteristics and additional information for all studied devices - part II.}
    \label{fig:summary2}
\end{figure} 

\newpage

\begin{figure}[h!]
\centering
     \includegraphics[width=\linewidth]{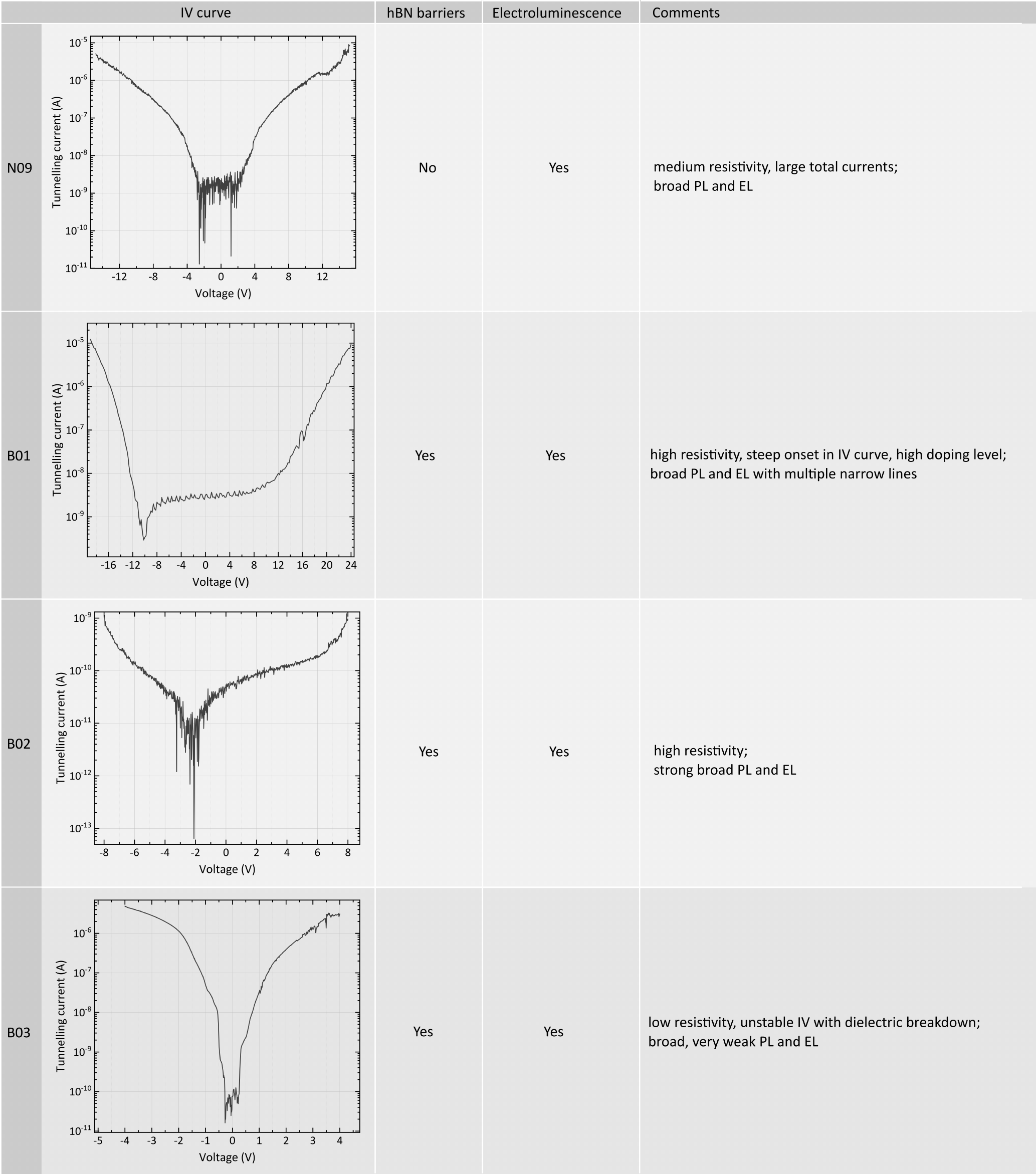}
    \caption{Current-voltage characteristics and additional information for all studied devices - part III.}
    \label{fig:summary3}
\end{figure}

\noindent Overall 14 samples were studied (12 summarized in Figs.~\ref{fig:summary1}-Figs.~\ref{fig:summary3} and 2 presented in the main text). Ten comprise hBN:C in between two graphene contacts (N01-N09), whereas the other four samples were fabricated with additional pristine hBN barriers (B01-B03). To ensure their functionality, each device was subjected to electrical testing via \textit{IV} characteristics. The \textit{IV} curves for all samples exhibit tunneling diode operation. However, the actual quality of device performance was found to vary from sample to sample. A common observation was that samples with hBN barriers exhibited higher tunneling thresholds than those without, highlighting the potential impact of the structure on device performance. While all samples 
displayed broad emission with varying intensity, only a few showed the presence of narrow lines, pointing to the essential role of the quality and architecture of the investigated devices.

\section{Device performance after dielectric breakdown}

\begin{figure}[h!]
\centering
    \includegraphics[width=\linewidth]{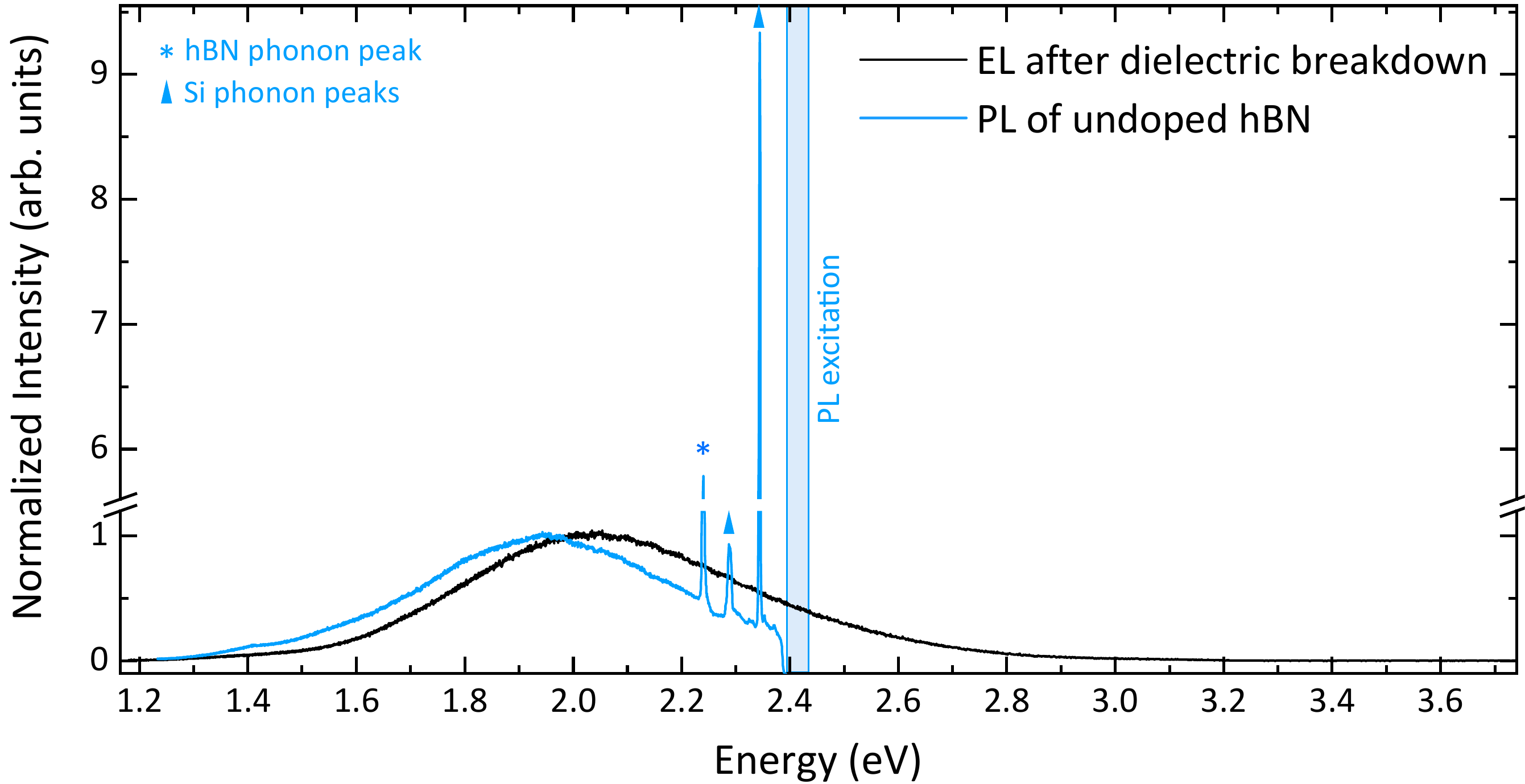}
    \caption{The comparison of the photoluminescence and electroluminescence spectra for the Gr/hBN/hBN:C/hBN/Gr device ater a dielectric breakdown. The photoluminescence spectrum was collected with 2.41~eV laser excitation. The laser was focused to a spot of 1~$\upmu$m diameter with a power of 1~mW. The electroluminescence was measured with -10~V bias voltage, which corresponded to I$_\text{T}$ = -27.4~$\upmu$A.}
    \label{fig:undoped}
\end{figure} 

\noindent During the course of the measurements, some of the samples underwent dielectric breakdown, in which their properties changed due to the application of high voltage. Surprisingly, some of the damaged samples still displayed broadband emission, indicating that their electroluminescent properties had not been entirely compromised. When compared with the photoluminescence of an undoped hBN sample, the emission curves exhibited similar characteristics, suggesting that during the breakdown process, the hBN:C material had been damaged, and the main recombination paths related to the defect states changed. There is a visible shift between the two curves of about 60~meV. This energy difference might be related to the dielectric environment in the studied samples.

\section*{Sample homogeneity}

\begin{figure}[h!]
\centering
    \includegraphics[width=\linewidth]{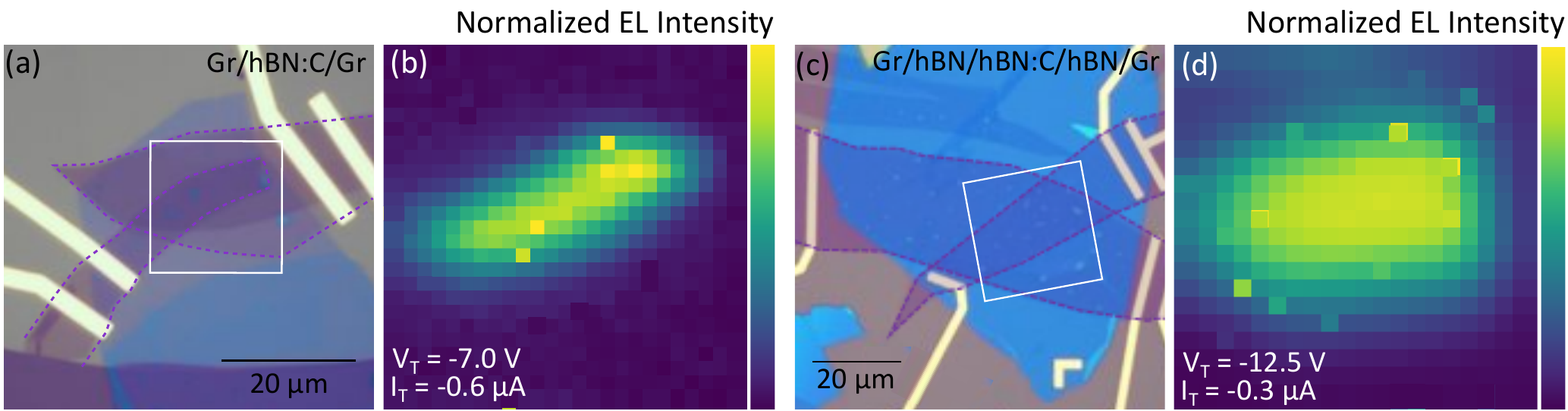}
    \caption{Optical microscope images of the representative devices for both examined architectures (a) Gr/hBN:C/Gr and  (c) Gr/hBN/hBN:C/hBN/Gr. Scanning electroluminescence images measured at bias V$_\text{T}$ of -7.0~V (b) and V$_\text{T}$ of -12.5~V (c). A homogenous signal at energy $\approx 2$~eV is observed where the hBN:C overlaps with the graphene electrodes.}
    \label{fig:maps}
\end{figure} 

\noindent Fig.~\ref{fig:maps}(a) shows an optical image of the Gr/hBN:C/Gr sample, with a designated area that was scanned, highlighted by a white square. A corresponding electroluminescence map of the sample is displayed in Fig.~\ref{fig:maps}(b) when a bias voltage of -7~V was applied. The same set of data is presented for Gr/hBN/hBN:C/hBN/Gr, with the optical image presented in Fig.~\ref{fig:maps}(c) and the electroluminescence map shown in Fig.~\ref{fig:maps}(d) when a bias voltage of -12.5~V was applied. In both cases the entire area emits light in a very similar manner, revealing that the samples are remarkably uniform in their electroluminescent properties. 

\section*{EL Intensity vs bias voltage.}

\begin{figure}[h!]
\centering
    \includegraphics[width=.95\linewidth]{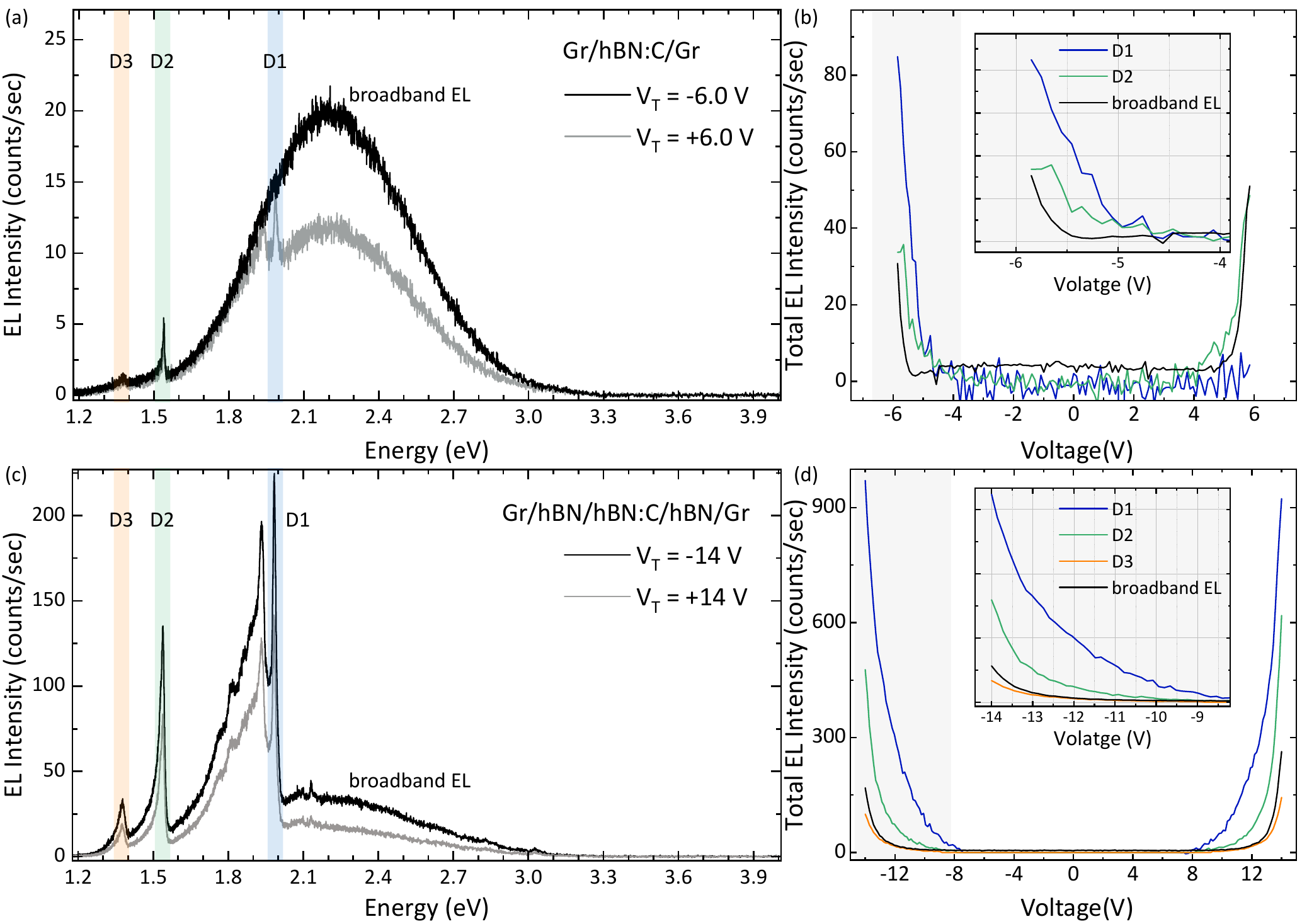}
    \caption{Electroluminescence spectra with positive and negative bias voltage (a/c) with the fitted EL intensity dependence on the tunnelling bias for selected lines (b/d) from devices without/with hBN barriers.}
    \label{fig:intensities}
\end{figure}

In the Fig.~\ref{fig:intensities}, we compare the activation threshold voltage for different features observed in the emission of two different structures: Gr/hBN:C/Gr and Gr/hBN/hBN:C/hBN/Gr. For the Gr/hBN:C/Gr structure, we observe that at lower bias voltages, the narrow lines related to defects D1 and D2 can be detected, while the broadband emission only emerges for bias voltages above 5.8~V. Additionally, the electroluminescence related to the D1 line shows asymmetric charge injection and is not effective at positive bias voltages. For the Gr/hBN/hBN:C/hBN/Gr structure, this observation is even more clear, with the emergence of the D1 emission line occurring at the lowest bias voltage ($\approx$ 7.3~V). As the voltage is increased, the other defects line is activated, with the D2 line observed above 9~V, followed by the detection of both the D3 line and broadband emission for voltages above 12.5~V. 

\section{Competition between the dielectric screening and Stark effect.}
\begin{wrapfigure}{r}{0.65\textwidth}
  \begin{center}
    \includegraphics[width=0.62\textwidth]{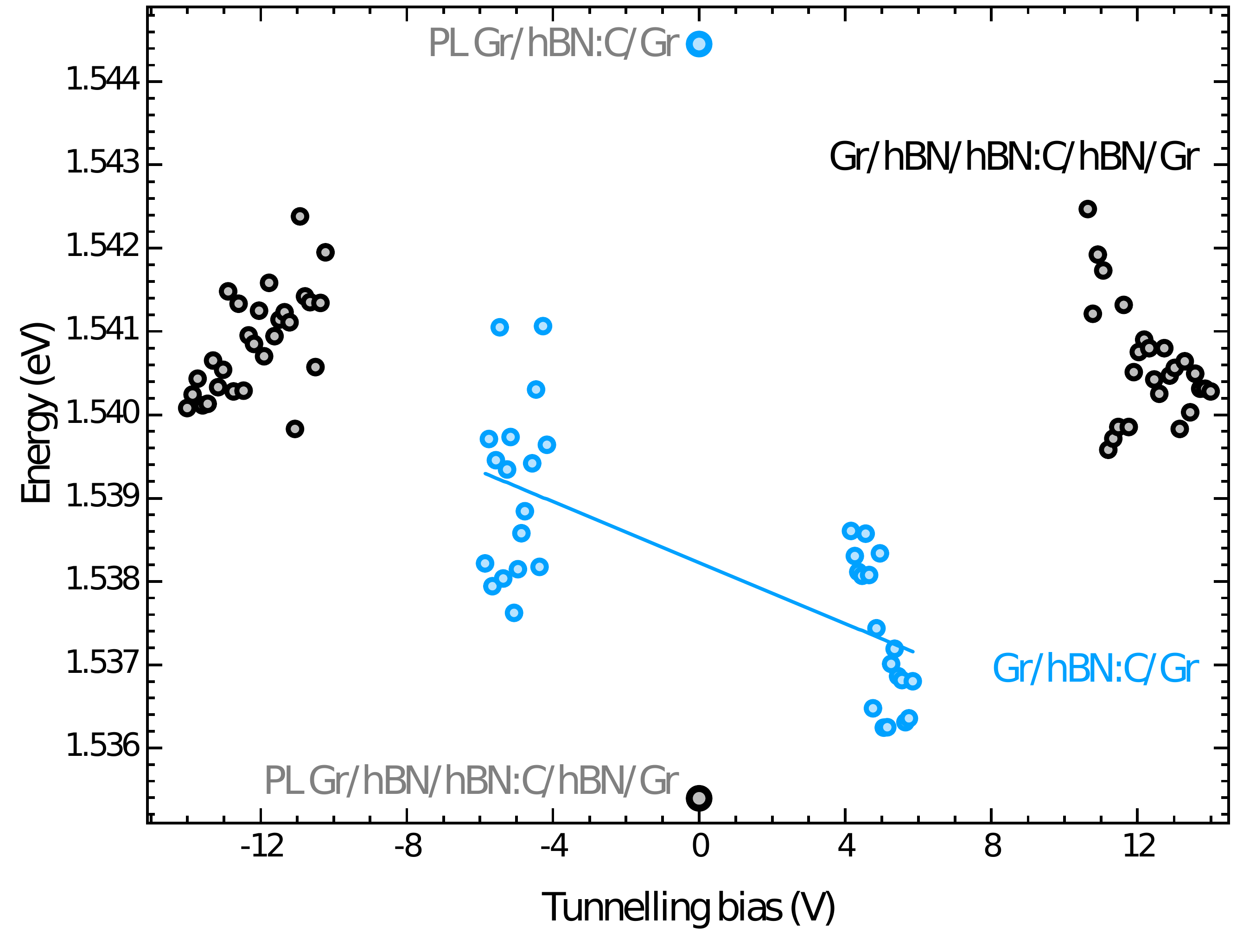}
  \end{center}
  \caption{Peak positions of the D2 defect as a
function of bias voltage for Gr/hBN:C/Gr (blue dots) and Gr/hBN/hBN:C/hBN/Gr (black dots) devices. The respective energies of the resonance in the photoluminescence spectra has been marked at 0~V.}
\end{wrapfigure}

Figure S7 illustrates the evolution of the energy of the D2 defect as a function of applied bias voltage. We focus our discussion on this feature as it is observable for both of the studied sample architectures. However, it is worth noting that in the case of the sample with a Gr/hBN/hBN:C/hBN/Gr structure, we observe the same behavior of the emission energy for all three types of defects (D1-D3) as the tunnelling bias is increased, as discussed in the main text. For the Gr/hBN:C/Gr device, there is a clear linear dependency of the emission energy with the increased tunnelling bias throughout the considered voltage range (-5.85 -- 5.85~V). This shift can be attributed to the confined Stark effect induced by the electric field, which changes the energy levels of the electrons in the system. A more complex behavior can be seen for the sample with the Gr/hBN/hBN:C/hBN/Gr structure. Initially, the energy of the emission increases with increasing bias voltage from negative values but then decreases with further positive bias. We interpret this result as the change in charge interaction influenced by the dielectric screening of the surrounding medium. The energy levels of the device are strongly affected by the dielectric screening, which can dominate the Stark effect. Such behavior is a common topic of research in quantum confined systems and is often used as a way of controlling excitonic complexes via external electrical fields.\cite{finley2004,abraham2021}

There is also a notable difference between the energy of the peak while exciting the system electrically and optically. For Gr/hBN:C/Gr the energy shift of the peak from PL and EL spectrum equals approximately +6.2~meV, while in the case of Gr/hBN/hBN:C/hBN/Gr device the difference is -7.2~meV. Such a disparity suggests that the sample may be in different states of excitation depending on its architecture. We can hypothesize the existence of several possible excitation complexes of defect states that engage additional electrons in the ground and/or excited state. These scenarios will be outlined in the next section.

\section{Single vs Many--body states}

\begin{figure}[h]
\centering
    \includegraphics[width=\linewidth]{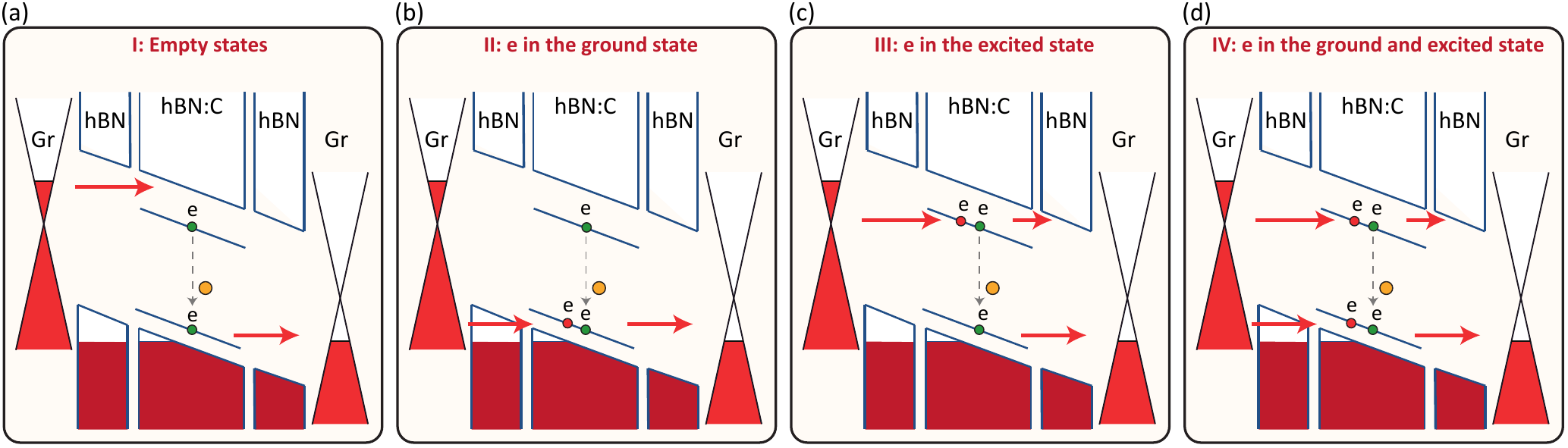}
    \caption{Schematic depiction of the additional many-body picture of charging and electroluminescence mechanism based on the band structure of the device and the Fermi level alignment with bias (a) considering an additional electron in the ground state (b), excited state (c) or both (d).}
    \label{fig:states}
\end{figure} 

\noindent We propose alternative many-body processes that can explain observed defect transitions in our electroluminescence results.  Specifically, we propose that additional electrons may be residing in the ground, excited, or both states, leading to these observed defect transitions and their particular behavior in different tunneling regimes. When additional electrons are involved, the exchange interactions between the electrons can affect the transition energies yielding the onset of electroluminescence. At high bias, the system can be in the ground electronic state, with an electron injected into the conduction band that relaxes to the underlying defect level. Similarly, a hole can be injected into the valence band, enabling the recombination of an electron occupying a defect level. Additionally, an electron can tunnel through an empty hBN band gap to the defect level in hBN:C. The specific electronic structure of the mid-gap levels, in combination with the position of the Fermi level in graphene relative to the hBN band edges forming the triangular tunnel barrier, determines whether an electron is inserted into the ground or excited defect state. This electron may further tunnel to the graphene layer, completing the charge transfer process. Overall, the behavior of the system under high bias is complex and depends on several factors, including the electronic properties of the defect levels and the energy barriers associated with tunneling through the hBN:C and graphene layers. Further investigation is necessary to fully understand these many-body processes and their contributions to the behavior of the system under different tunnelling regimes.

\end{document}